\documentclass[a4paper,11pt]{article}
\usepackage{pos}
\usepackage{lineno}
\pdfoutput=1 
\title{Quarkonia production as a function of charged-particle multiplicity in pp collisions at $\sqrt{s}=13$~TeV with ALICE }
\ShortTitle{Quarkonia production as a function of multiplicity}

\author*[a,b]{Yanchun Ding}
\author {on behalf of the ALICE Collaboration}

\affiliation[a]{Key Laboratory of Quark and Lepton Physics (MOE) and Institute of Particle Physics, Central China Normal University, Wuhan 430079 China\\}

\affiliation[b]{Institut de Physique des 2 Infinis de Lyon, Université Claude Bernard Lyon 1,\\
4 rue Enrico Fermi, 69622 Villeurbanne Cedex, France}
\emailAdd{yanchun.ding@cern.ch}

\abstract{
In pp collisions at LHC energies, heavy quarks are produced in initial hard scatterings and then these quarks hadronize in either open heavy-flavor hadrons or quarkonia (e.g. J/$\psi$, $\psi(2S)$, $\Upsilon$). The study of quarkonium production as a function of charged-particle multiplicity links soft and hard processes and allows one to study their interplay. While a linear increase of quarkonium production as a function of charged-particle multiplicity can be reasonably well understood in the context of multi-parton interactions, the observation of deviations with respect to a linear increase requires a more detailed description of the collision.

In this contribution, we will present the latest ALICE measurements at forward rapidity for J/$\psi$ and $\Upsilon$ production as a function of charged-particle multiplicity in pp collisions at $\sqrt{s}=13$ TeV. The first measurement of the double ratios of relative yield of $\Upsilon(2S)$ over $\Upsilon(1S)$ and J/$\psi$ over $\Upsilon(1S)$ as a function of charged-particle multiplicity will also be shown.
}

\FullConference{%
  The Eighth Annual Conference on Large Hadron Collider Physics-LHCP2020 \\
  25-30 May, 2020\\
  online}

\begin{document}
\maketitle 
\section{Introduction} 
The event-multiplicity dependent production of quarkonium and open heavy-flavor hadrons in small collision systems such as pp and p--Pb is widely studied at the LHC, because it has the potential to give new insights on processes at the parton level and on the interplay between the hard and soft mechanisms in particle production. ALICE has studied the multiplicity dependence in pp collisions at $\sqrt{s} = 13$~TeV of inclusive J/$\psi$ production at mid-rapidity \cite{Ref:Jpsi-Mid}, which shows a stronger than linear increasing trend. It is  compared with various theoretical models, such as the coherent production model \cite{Ref:coherent}, the CGC model \cite{Ref:CGC}, the 3-Pomeron CGC model \cite{Ref:3-Pomeron CGC}, and PYTHIA 8.2
predictions \cite{Ref:Pythia8_1,Ref:Pythia8_2}. With similar motivations, the recent multiplicity dependence of $\Upsilon$ production at forward rapidity has been studied, aimed to improve the understanding of the underlying production mechanisms.

\section{Analysis strategy}
\begin{figure}[!htpb]
\begin{center}
\includegraphics[width=.48\textwidth]{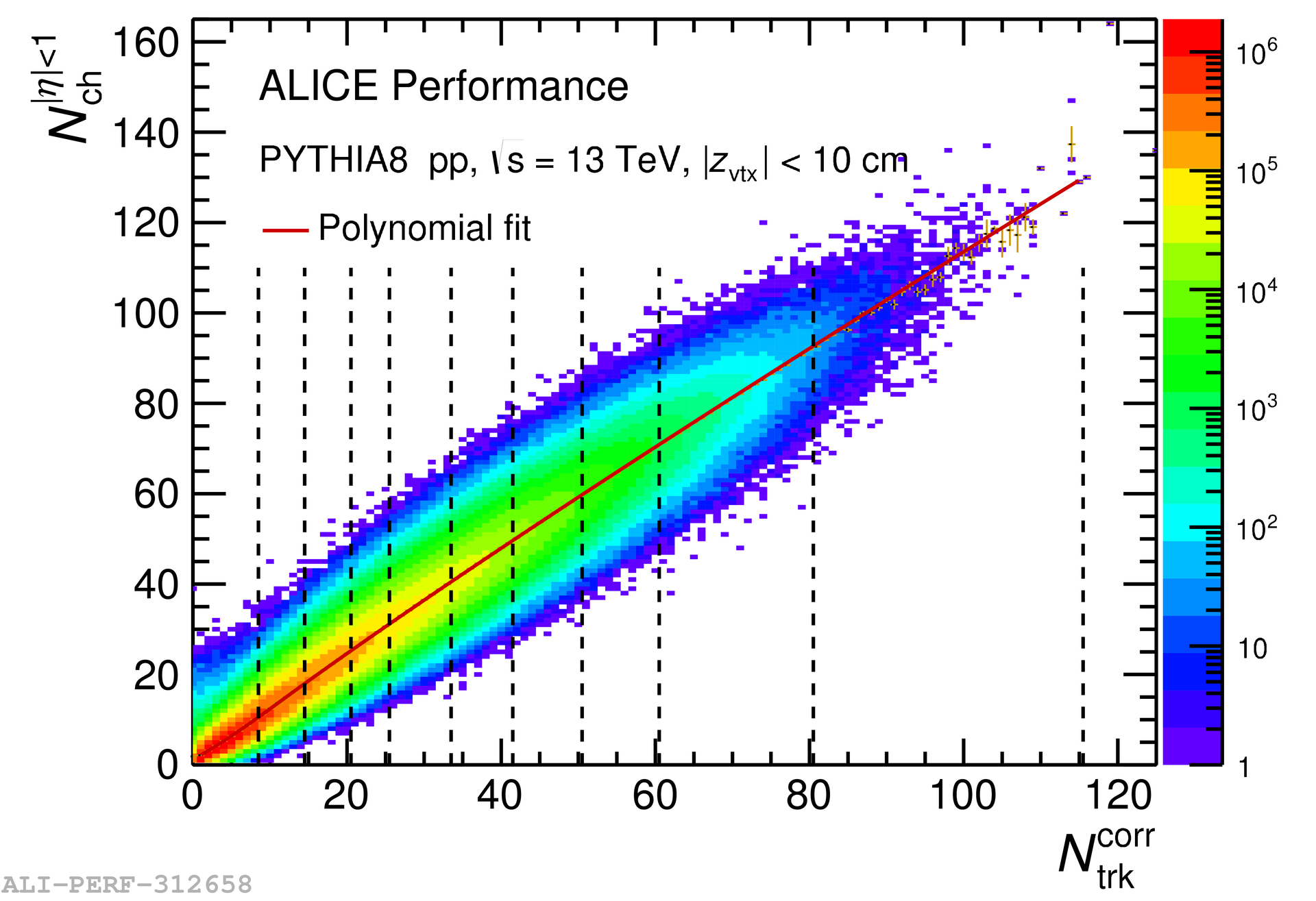}
\end{center}
\caption{Number of charged particles $N_{\mathrm {ch}}$ as a function of trakclets $N_{\mathrm {trk}}^{\mathrm {corr}}$ as determined by a Monte Carlo simulation using PYTHIA 8 simulation with superimposed the best fit with a polynomial function.}
\label{Fig:Ntrk-Nch}
\end{figure}

In this analysis, tracklets, i.e. track segments reconstructed in the ALICE Silicon Pixel detector (SPD) \cite{Ref:detector} with pseudorapidity $|\eta| < 1$, are used for the charged-particle multiplicity estimation. The first step of the multiplicity calibration is to correct for the detector inefficiency along the interaction vertex ($z_{\mathrm {vtx}}$), by equalizing the number of tracklets variation as a function of $z_{\mathrm{vtx}}$ on an event by event basis. Then, using a Monte Carlo simulation based on the PYTHIA 8.2 \cite{Ref:Pythia8_1} and EPOS-LHC \cite{Ref:epos} event generators, the correlation between the tracklet multiplicity (after the $z_{\mathrm {vtx}}$-correction), $N_{\mathrm {trk}}^{\mathrm {corr}}$, and the generated primary charged particles $N_{\mathrm {ch}}$ is determined, as shown in Fig.~\ref{Fig:Ntrk-Nch}. Finally, the self-normalized multiplicity is defined as the ratio between the charged-particle density (d$N_{\mathrm {ch}}$/d$\eta$) in a given multiplicity interval to the integrated one.

\begin{figure}[!htpb]
\begin{center}
\includegraphics[width=.48\textwidth]{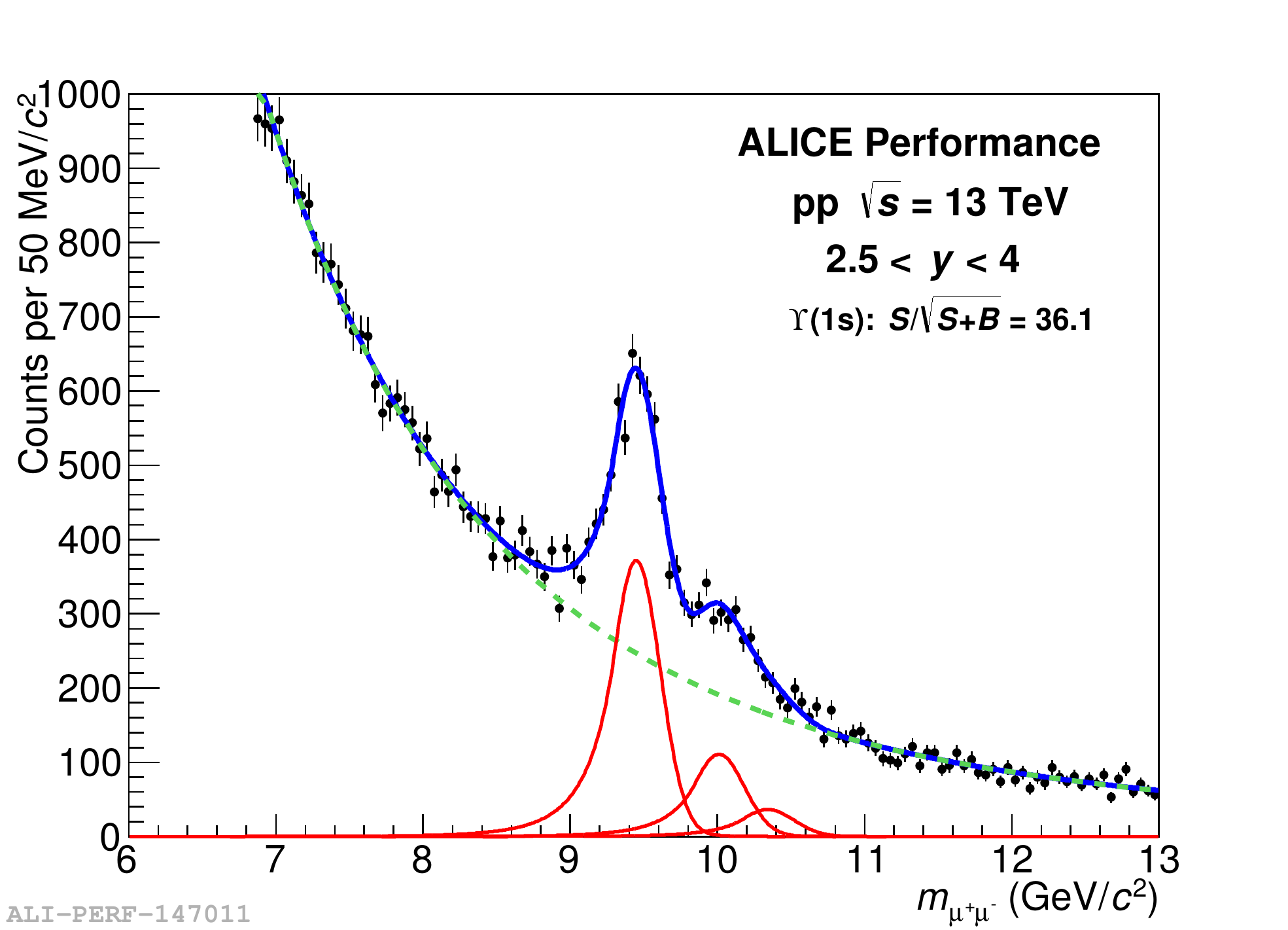}
\includegraphics[width=.48\textwidth]{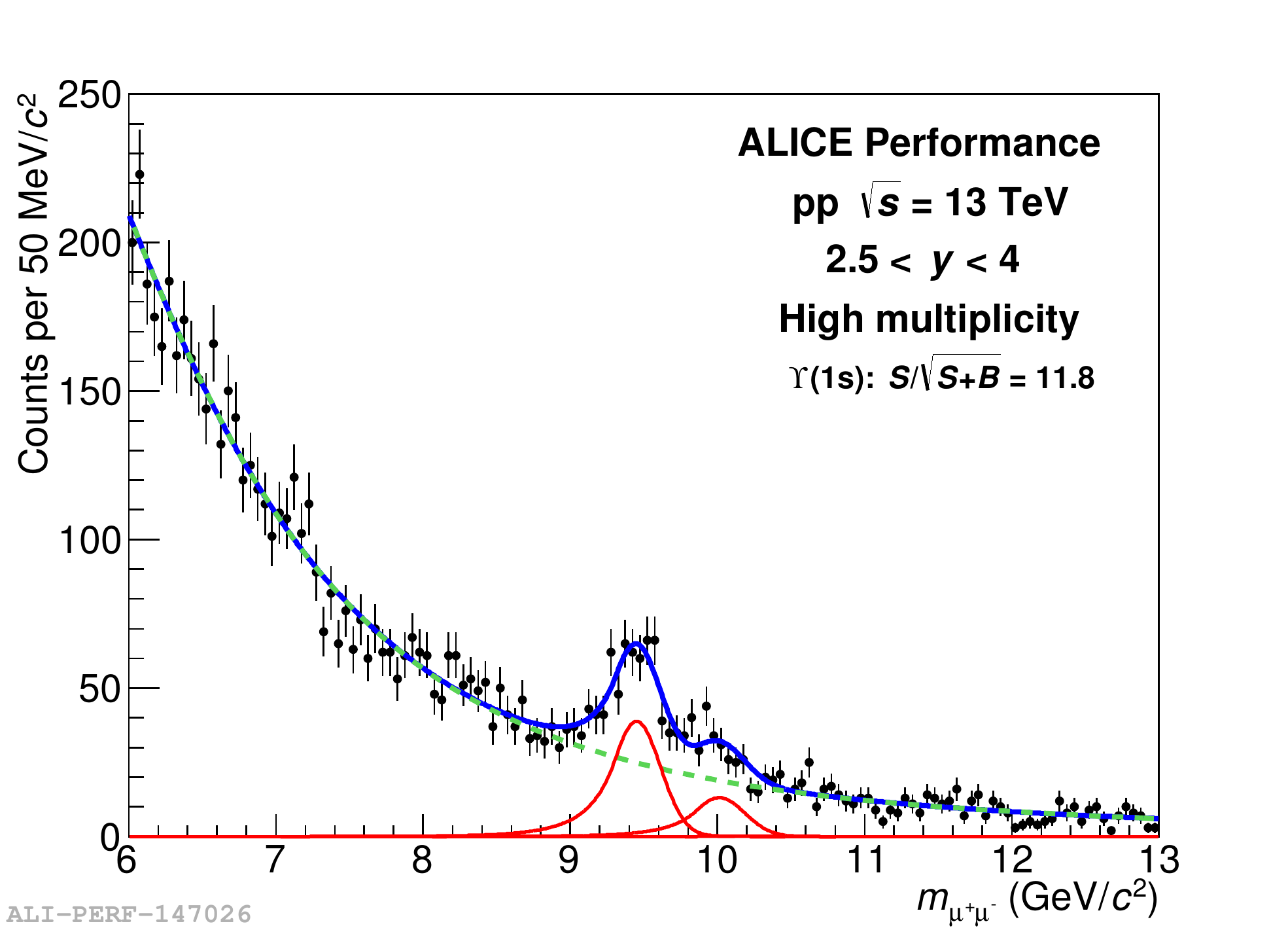}
\end{center}
\caption{Di-muon invariant mass distribution for integrated over the multiplicity (left) and for high multiplicity pp
collisions, corresponding to the $N_{\mathrm {trk}}^{\mathrm {corr}}$ interval bin $[34, 41]$ (right).}
\label{Fig:Upsilon}
\end{figure}

The $\Upsilon$ is reconstructed in the rapidity range $-4.0 < y < -2.5$ via the dimuon decay channel using the forward muon spectrometer~\cite{Ref:detector}. The dimuon invariant mass distribution in the region relevant for $\Upsilon$ measurement is shown in Fig.~\ref{Fig:Upsilon} for the analyzed data sample. A log-likelihood fit is applied to extract the signal. The signal shape is described by a Double Crystal Ball (DCB) function and the background shape is described by a
Variable Width Gaussian (VWG) function ~\cite{Ref:fitFunc}. 

\section{Results and conclusions}
\begin{figure}[!htpb]
\begin{center}
\includegraphics[width=.48\textwidth]{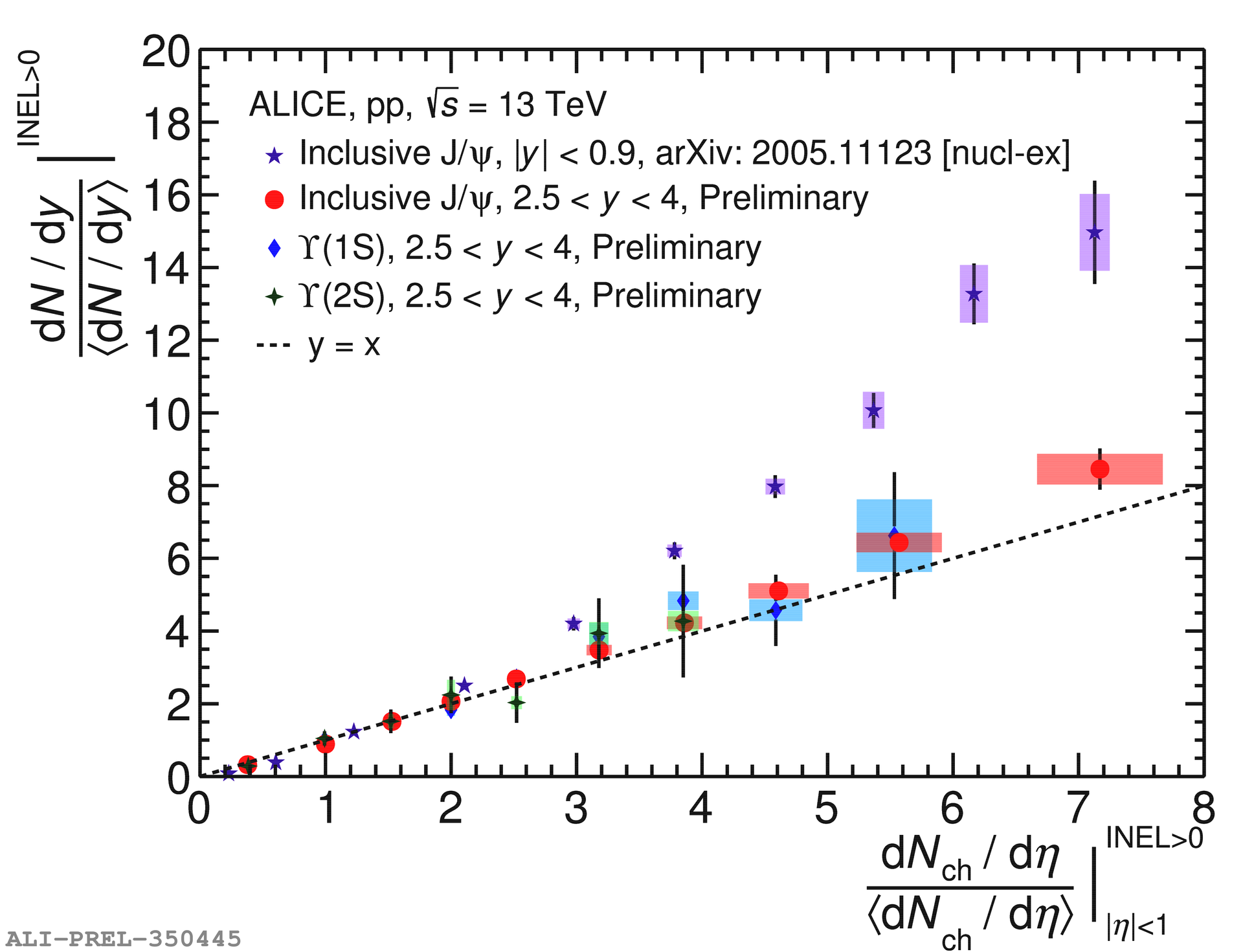}
\end{center}
\caption{Self-normalized yield of $\Upsilon$ and J/$\psi$ as a function of normalized charged particle multiplicity.}
\label{Fig:QuarkoniaVsMult}
\end{figure}

\begin{figure}[!htpb]
\begin{center}
\includegraphics[width=.48\textwidth]{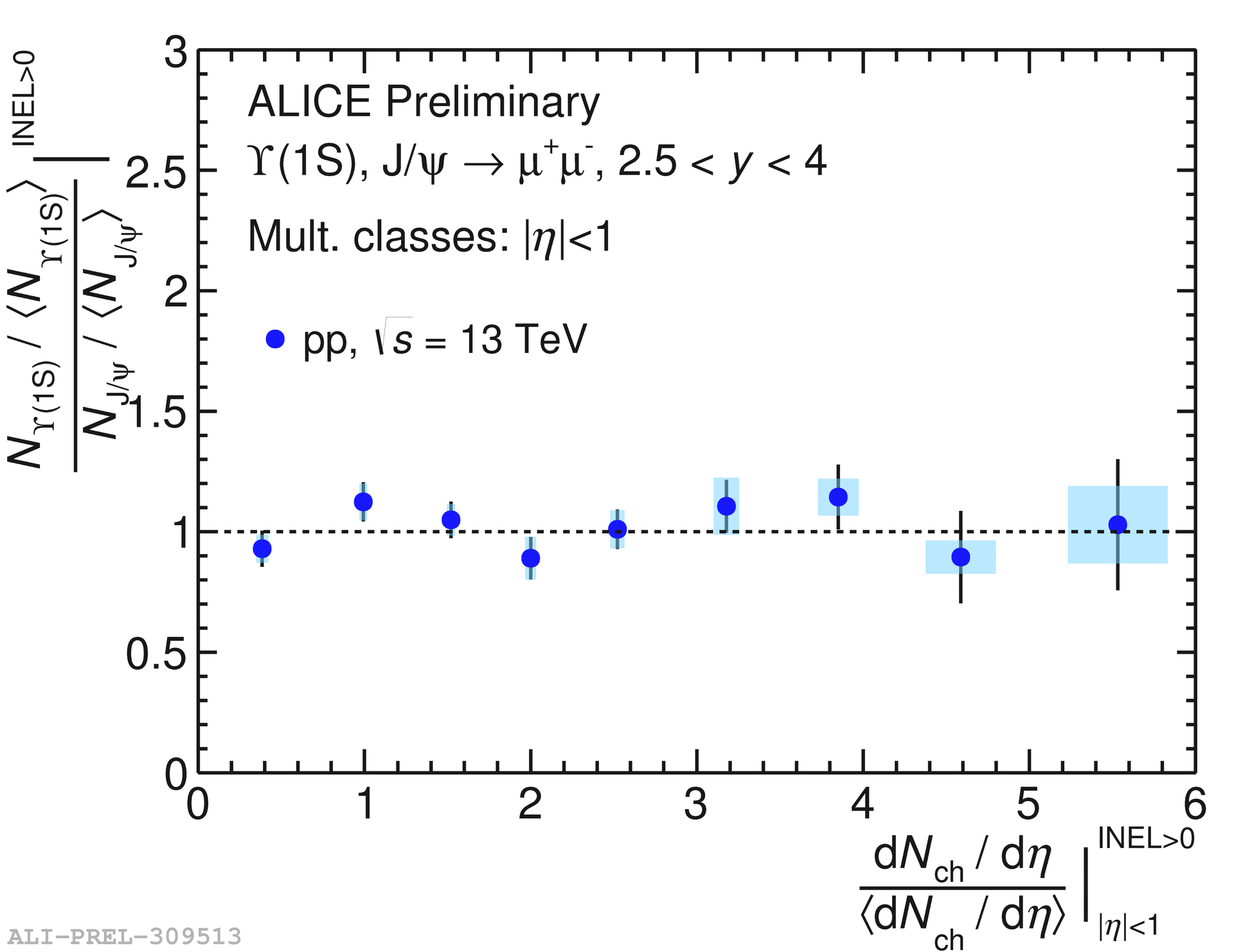}
\includegraphics[width=.48\textwidth]{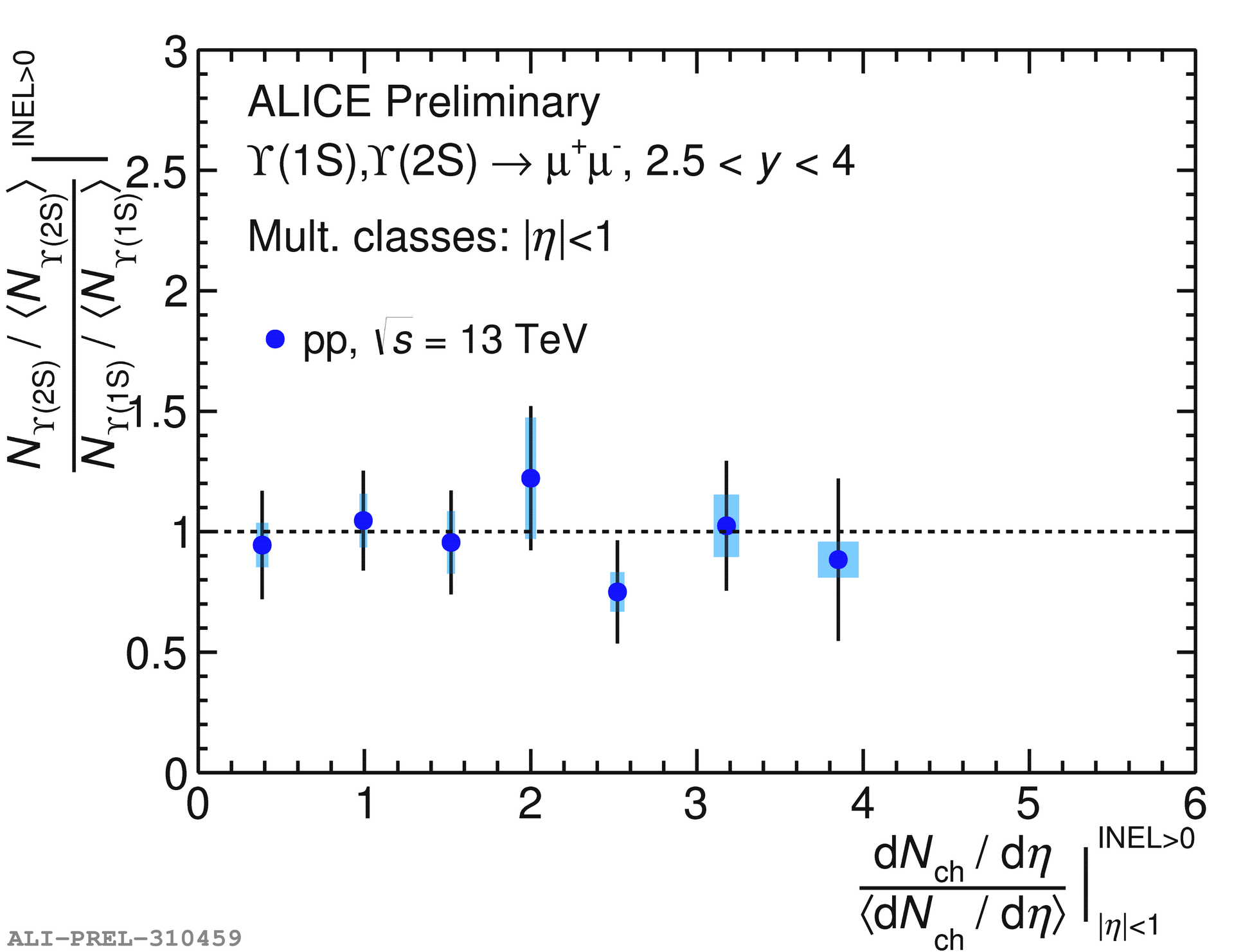}
\end{center}
\caption{Left: Double yield ratio of $\Upsilon(1S)$ over J/$\psi$ as a function of normalized charged-particle multiplicity; Right: Double yield ratio of $\Upsilon(2S)$ over $\Upsilon(1S)$ as a function of normalized charged-particle multiplicity.}
\label{Fig:DoubleRatioVsMult}
\end{figure}

The self-normalized yield of $\Upsilon$ is defined as the $\Upsilon$ yield in a given multiplicity interval to the multiplicity-integrated yield.
As shown in Fig.~\ref{Fig:QuarkoniaVsMult}, an
approximately linear increasing behavior is observed for $\Upsilon(1S)$ (blue points), $\Upsilon(2S)$ (green points) and J/$\psi$ (red points) at forward rapidity. However, a faster than linear increase is presented for J/$\psi$ (purple points) at mid-rapidity,
when there might be a correlation between the signal ($|y| < 0.9$) and the multiplicity estimator ($|\eta| < 1$). The different trends observed with the introduction of a rapidity gap might be explained as the result of largely eliminating the auto-correlation effects~\cite{Ref:Pythia8_2}. The self-normalized yield ratios of $\Upsilon(2S)$ over $\Upsilon(1S)$ and $\Upsilon(1S)$ over J/$\psi$, as shown in Fig.~\ref{Fig:DoubleRatioVsMult}, are independent on multiplicity and compatible with unity within uncertainties. It reveals that there is no dependence on resonance mass and quark component within uncertainties.

In this contribution, the first results of the $\Upsilon(1S)$ and $\Upsilon(2S)$ production as a function of charged-particle multiplicity have been presented. A different behavior is observed compared with the J/$\psi$ at mid-rapidity. In addition, the self-normalized yield ratios of $\Upsilon(1S)$ over J/$\psi$ and $\Upsilon(2S)$ over $\Upsilon(1S)$ as a function of multiplicity are found to be compatible with unity within uncertainties. 


\end{document}